\documentstyle[sprocl]{article}

\newcommand{\beq}{\begin{eqnarray}}
\newcommand{\eeq}{\end{eqnarray}}
\newcommand{\drawsquare}[2]{\hbox{%
\rule{#2pt}{#1pt}\hskip-#2pt
\rule{#1pt}{#2pt}\hskip-#1pt
\rule[#1pt]{#1pt}{#2pt}}\rule[#1pt]{#2pt}{#2pt}\hskip-#2pt
\rule{#2pt}{#1pt}}

\newcommand{\Yfund}{\raisebox{-.5pt}{\drawsquare{6.5}{0.4}}}
\newcommand{\Ysymm}{\raisebox{-.5pt}{\drawsquare{6.5}{0.4}}\hskip-0.4pt%
        \raisebox{-.5pt}{\drawsquare{6.5}{0.4}}}
\newcommand{\Ythrees}{\raisebox{-.5pt}{\drawsquare{6.5}{0.4}}\hskip-0.4pt%
          \raisebox{-.5pt}{\drawsquare{6.5}{0.4}}\hskip-0.4pt%
          \raisebox{-.5pt}{\drawsquare{6.5}{0.4}}}
\newcommand{\Yasymm}{\raisebox{-3.5pt}{\drawsquare{6.5}{0.4}}\hskip-6.9pt%
        \raisebox{3pt}{\drawsquare{6.5}{0.4}}}
\newcommand{\Ythreea}{\raisebox{-3.5pt}{\drawsquare{6.5}{0.4}}\hskip-6.9pt%
        \raisebox{3pt}{\drawsquare{6.5}{0.4}}\hskip-6.9pt
        \raisebox{9.5pt}{\drawsquare{6.5}{0.4}}}
\newcommand{\Yfoura}{\raisebox{-3.5pt}{\drawsquare{6.5}{0.4}}\hskip-6.9pt%
        \raisebox{3pt}{\drawsquare{6.5}{0.4}}\hskip-6.9pt
        \raisebox{9.5pt}{\drawsquare{6.5}{0.4}}\hskip-6.9pt
        \raisebox{16pt}{\drawsquare{6.5}{0.4}}}
\newcommand{\Yadjoint}{\raisebox{-3.5pt}{\drawsquare{6.5}{0.4}}\hskip-6.9pt%
        \raisebox{3pt}{\drawsquare{6.5}{0.4}}\hskip-0.4pt
        \raisebox{3pt}{\drawsquare{6.5}{0.4}}}
\newcommand{\jref}[4]{{\it #1} {\bf #2}, #3 (#4)}

\newcommand{\MPLA}[3]{\jref{Mod.\ Phys.\ Lett.}{A#1}{#2}{#3}}

\newcommand{\NPB}[3]{\jref{Nucl.\ Phys.}{B#1}{#2}{#3}}

\newcommand{\PLB}[3]{\jref{Phys.\ Lett.}{#1B}{#2}{#3}}

\newcommand{\PRD}[3]{\jref{Phys.\ Rev.}{D#1}{#2}{#3}}

\newcommand{\PRL}[3]{\jref{Phys.\ Rev.\ Lett.}{#1}{#2}{#3}}

\def\be{\begin{equation}}
\def\ee{\end{equation}}
\def\bea{\begin{eqnarray}}
\def\eea{\end{eqnarray}}

\begin{document}
\begin{center}
hep-th/9807222     \hfill    LBNL-42126 \\
\end{center}
\vspace*{0.5cm}

\title{THE CONFINING $N=1$ SUPERSYMMETRIC GAUGE THEORIES: 
A REVIEW\footnote{Talk 
presented at the 3rd workshop on Continuous Advances in QCD, 
Minneapolis, MN, 16-19 April 1998.}}

\author{CSABA CS\'AKI\footnote{Research fellow, Miller Institute for 
Basic Research in Science.}}

\address{Theoretical Physics Group\\
     Ernest Orlando Lawrence Berkeley National Laboratory\\
     University of California, Berkeley, California 94720\\
and \\
Department of Physics\\
     University of California, Berkeley, California 94720}


\maketitle\abstracts{We give a classification and overview of the confining
$N=1$ supersymmetric gauge theories. For simplicity we consider only theories
based on simple gauge groups and no tree-level superpotential. Classification
of these theories can be done according to whether or not there is a
superpotential generated for the confined degrees of freedom. The 
theories with the superpotential include s-confining theories and also
theories where the gauge fields participate in the confining spectrum,
while theories with no superpotential include theories with a quantum 
deformed moduli space and theories with an affine moduli space.}

\section{Introduction}

In this paper, we give an overview of the confining $N=1$ supersymmetric
gauge theories. Before jumping into the details of the classification of
such models, one has to answer the question of what we mean by a confining
theory. The definition we will be using throughout this paper is the 
following: {\it we call a theory confining, if there is a low-energy 
description purely in terms of composite gauge singlets} (that is, the 
low-energy effective theory is a Wess-Zumino model for the gauge singlet 
fields, there are no massless gauge degrees of freedom in the IR theory).
This broad definition of confinement does not automatically 
imply that there would be an area law for the Wilson loop, or a linear 
potential between external test charges. The reason is that in some cases
(when there are massless dynamical fields in a faithful representation of 
the gauge group), the external charges can be screened, and instead
of a linear potential there will be no potential at all. In this case
there is no phase boundary between the Higgs and the confining phases,
and there is 
no invariant distinction between these two phases. This is what will
actually happen in most of the examples reviewed below. Keeping this
broad definition of confinement in mind, we are ready to discuss
the classification of these models.

For simplicity, we will consider only theories based on simple gauge 
groups and no tree-level superpotential. Then the confining theories
can be classified into two broad categories, according to whether 
or not there is a superpotential generated for the composite fields.
These two categories can be further refined: for the case of theories with
a confining superpotential, one can distinguish between theories where
the composites contain only chiral superfields (these are the
s-confining theories), or theories where the gauge field $W_{\alpha}$ also 
participates in forming the composites. In the case of theories with no
superpotential, one can distinguish between theories where there are 
classical constraints relating the composites and theories where there are
no such constraints. These categories will be discussed in detail below.
A final category which we will not discuss in detail is when the low-energy
effective theory is empty, that is there is a mass gap, and no 
massless chiral superfields are present. This is the case for example for 
$N=1$ pure Yang-Mills theories. However, we expect such theories 
to be very rare for the following reason. If there is an exact continuous
global symmetry present in the theory, then it is either spontaneously
broken or not. If it is spontaneously broken, we expect massless
Goldstone-bosons, if it is not broken, then the 't Hooft anomaly matching
conditions have to be satisfied, implying the presence of massless fermions.
Thus we expect that only theories like pure $N=1$ Yang-Mills, with no
continuous global symmetries to exhibit such behavior. Finally, a warning:
the four categories to be explained below contain all confining theories 
known up today. However, it is possible, that there might be a lot 
more confining theories around, which might not fit into the above 
classification scheme.

\section{Theories with a Non-vanishing Confining Superpotential}

\subsection{The S-confining Theories}

S-confining theories are defined as follows~\cite{CSS}:
 
- there is a non-vanishing superpotential for the confined degrees of
freedom (non-singular at the origin)

- the composites involve only chiral superfields 

- the description in terms of gauge invariant composites is valid everywhere
on the moduli space.

\noindent The first example of an s-confining theory has been found by
Seiberg~\cite{Seiberg}. We will use this example ($SU(N)$ theory with
$F=N+1$ flavors) to explain the most important properties of such theories.
The field content and global symmetries of the theory, together with the
confining spectrum is given below.
\[
\begin{array}{c|c|cccc}
& SU(N) & SU(N+1) & SU(N+1) & U(1)_B & U(1)_R \\ \hline
Q & \Yfund & \Yfund & 1 & 1 & \frac{1}{N+1} \\
\bar{Q} & \overline{\Yfund} & 1 & \Yfund & -1 & \frac{1}{N+1} \\ \hline \hline
M=(\bar{Q}Q) & & \Yfund & \Yfund & 0 & \frac{2}{N+1} \\
B=(Q^N) &  & \overline{\Yfund} & 1 & N & \frac{N}{N+1} \\
\bar{B} =(\bar{Q}^N) & & 1 & \overline{\Yfund} & -N & \frac{N}{N+1}
\end{array} \]
The confining superpotential is 
\beq 
\frac{1}{\Lambda^{2N-1}}({\rm det}\, M -\bar{B}MB)
\eeq 
There is ample of evidence that this is indeed the correct low-energy
description of the original $SU(N)$ theory~\cite{Seiberg}. First of all, the 
confined degrees of freedom $M,B$ and $\bar{B}$ satisfy the
't Hooft anomaly matching conditions. Second, the classical limit is 
correctly reproduced by the superpotential, since the equations of
motion result exactly in the classical constraints of the theory. 
Finally, integrating out flavors results in the correct descriptions of the
theories with less flavors. Subsequently, several other s-confining theories 
have been
found~\cite{Pouliot,Sp}. The natural question to ask is how to find all
other s-confining theories. We will answer this question below.

The most severe constraint on s-confining theories comes from the
requirement that there is a non-vanishing confining superpotential.
Global symmetries fix this superpotential to be of the form~\cite{CSS}
\beq
W \propto \Lambda^3 \left[ \prod_i \frac{ \Phi_i^{\mu_i}}{\Lambda^{\mu_i}}
\right]^{\frac{2}{\sum_i \mu_i-G}},
\label{suppot}
\eeq
where $\Phi_i$ are the underlying chiral superfields (not the composites),
$\mu_i$ is the Dynkin index with respect to the gauge group of the
$i^{th}$ chiral superfield given by ${\rm Tr}\, T_A^iT_B^i=\mu_i \delta_{AB}$,
where the $T$'s are the generators of the gauge group in the $i^{th}$
representation, and $G$ is the Dynkin index of the adjoint. For example in
the case of Seiberg's example $SU(N)$ with $N+1$ flavors 
$\Phi_i=Q_i,\bar{Q}_i$ ($i$ is the flavor index $i=1,\ldots ,N+1$),
$\mu_i=1$, $G=2N$, thus the superpotential has the form 
$Q^{N+1}\bar{Q}^{N+1}/\Lambda^{2N-1}$, which can be written in terms of the
confined degrees of freedom either as ${\rm det}\, M$ or $\bar{B}MB$. 

Examining the form of (\ref{suppot}) one can observe, that the confining
superpotential is singular at the origin unless the overall
exponent is an integer, implying the index constraint
\beq 
\sum_i \mu_i-G= 2 \; \mbox{or}\;  1.
\eeq
This is a very  severe constraint on the matter content of a given 
theory. In fact, it restricts the candidates for s-confining theories
to a finite set. This set of theories for the case of $SU(N)$ groups
is given in Table~\ref{tab:s-conf}.
 \begin{table}
\caption{All $SU$ theories satisfying $\sum_i \mu_i - G = 2$.
This list is finite because the indices of higher index tensor
representations grow very rapidly with the size of the gauge group.
We give the gauge group in the first column, and the field content 
in the second column. In the third column, 
we indicate which theories are s-confining.
For the theories which do not s-confine we give the flows to non s-confining
theories or indicate that there is a Coulomb branch on the moduli space.
\label{tab:s-conf}}
\vspace*{1cm}
\begin{center}
\begin{tabular}{|l|l|l|} \hline
$SU(N)$ & $(N+1) (\Yfund + \overline{\Yfund})$ & s-confining \\
$SU(N)$ & $\Yasymm + N\, \overline{\Yfund} + 4\, \Yfund $ & s-confining \\
$SU(N)$ & $\Yasymm + \overline{\Yasymm} + 3 (\Yfund + \overline{\Yfund})$ &
  s-confining \\
$SU(N)$ & Adj  $+\Yfund + \overline{\Yfund}$ & Coulomb branch \\ \hline
$SU(4)$ & Adj $+ \Yasymm $ & Coulomb branch \\
$SU(4)$ & $3\, \Yasymm + 2 (\Yfund + \overline{\Yfund})$ &
   $SU(2)$: $8\, \Yfund$ \\
$SU(4)$ & $ 4\, \Yasymm + \Yfund + \overline{\Yfund}$ &
   $SU(2)$: $\Ysymm + 4\, \Yfund$  \\
$SU(4)$ & $ 5\, \Yasymm $ & Coulomb branch \\
$SU(5)$ & $ 3 (\Yasymm + \overline{\Yfund}) $ & s-confining \\
$SU(5)$ & $ 2\, \Yasymm + 2\, \Yfund + 4\, \overline{\Yfund}$ & s-confining \\
$SU(5)$ & $ 2 (\Yasymm + \overline{\Yasymm})$ & 
   $Sp(4)$: $3\, \Yasymm + 2\, \Yfund$ \\
$SU(5)$ & $2\, \Yasymm + \overline{\Yasymm} + 2\, \overline{\Yfund} +
  \Yfund$ &  $SU(4)$: $3\, \Yasymm + 
  2 (\Yfund + \overline{\Yfund})$ \\
$SU(6)$ & $2\, \Yasymm + 5\, \overline{\Yfund} + \Yfund$ &
  s-confining \\
$SU(6)$ & $ 2\, \Yasymm + \overline{\Yasymm} + 2\, \overline{\Yfund}$ &
   $SU(4)$: $3\, \Yasymm + 2 (\Yfund + \overline{\Yfund})$ \\
$SU(6)$ & $\Ythreea + 4 (\Yfund + \overline{\Yfund})$ & s-confining \\
$SU(6)$ & $\Ythreea + \Yasymm +  3\, \overline{\Yfund} + \Yfund$
  &  $SU(5)$: $2\, \Yasymm +
  \overline{\Yasymm} + 2\, \overline{\Yfund} + \Yfund $ \\
$SU(6)$ & $\Ythreea + \Yasymm + \overline{\Yasymm}$ & 
  $Sp(6)$: $\Ythreea + \Yasymm + \Yfund$ \\
$SU(6)$ & $2\, \Ythreea + \Yfund + \overline{\Yfund}$ &  $SU(5)$:
  $  2 (\Yasymm + \overline{\Yasymm})$ \\
$SU(7)$ & $2 (\Yasymm + 3\, \overline{\Yfund})$ & s-confining \\
$SU(7)$ & $ \Ythreea + 4\, \overline{\Yfund} + 2\, \Yfund$ &
   $SU(6)$: $\Ythreea + \Yasymm + 3\, \overline{\Yfund}+ \Yfund$ \\
$SU(7)$ & $\Ythreea + \overline{\Yasymm} + \Yfund$ & $Sp(6)$: $\Ythreea + 
   \Yasymm + \Yfund$ \\
  \hline \end{tabular}
\end{center}
\end{table}
In order to find out which of those theories listed in Table~\ref{tab:s-conf}
are actually s-confining, we note one more necessary condition the
s-confining theories have to satisfy: an s-confining theory flows only
to s-confining theories. The reason behind this is simple. An s-confining
theory is described by a set of gauge invariant operators. Going along
a flat direction in this language just means giving VEV's to some gauge
invariant fields, thus the resulting theory also has to be describable in terms
of a theory of gauge invariants. Using this condition one can go ahead and
check the various flows of the candidate theories listed in 
Table~\ref{tab:s-conf}. The theories where a flow results in a non-s-confining
theory can be excluded. For the remaining examples one can explicitly 
find the confined spectrum and show that the consistency conditions
are all satisfied. This way one can find all s-confining theories based
on simple groups. The results for $SU(N)$ theories are listed in 
Table~\ref{tab:s-conf}. Here we give just one more simple s-confining example,
which is based on $SU(5)$ with three antisymmetric tensors and three 
antifundamentals. The detailed description of the remaining $SU(N)$ theories
together with the theories based on other groups can be found in~\cite{CSS}.

\begin{displaymath}
\begin{array}{c|c|cccc}
& SU(5) & SU(3) & SU(3) & U(1) & U(1)_R \\ \hline
A & \Yasymm  &  \Yfund  & 1 & 1 & 0 \\
\bar{Q} & \overline{\Yfund} & 1 & \Yfund & -3 & \frac{2}{3} 
\\ \hline \hline
A\bar{Q}^2 & & \Yfund & \overline{\Yfund} & -5 & \frac{4}{3} \\
A^3\bar{Q} & & \Yadjoint & \Yfund & 0 & \frac{2}{3} \\
A^5 & & \Ysymm & 1 & 5 & 0\end{array}
\end{displaymath}

\[ W_{dyn}=\frac{1}{\Lambda^9} \Big[ (A^5)(A^3\overline{Q})(A\overline{Q}^2)+
(A^3\overline{Q})^3 \Big] \]

\subsection{Composites Contain $W_{\alpha}$}

This category does not have its own name, since there is only one known
example~\cite{IntSeib}. This example is based on an $SO(N)$ theory
with $N-3$ vectors. Intriligator and Seiberg argued, that there is a
branch on which the theory confines with the following spectrum:
\[ \begin{array}{c|c|ccc}
& SO(N) & SU(N-3) & U(1)_R & Z_{2N-6} \\ \hline
Q & \Yfund & \Yfund & \frac{-1}{N-3} & 1 \\ \hline \hline 
M =(Q^2)& & \Ysymm & \frac{-2}{N-3} & 2 \\
b =(W_{\alpha}^2Q^{N-4})&  
& \overline{\Yfund} & 1+\frac{1}{N-3} & N-4 \end{array}\]
The confining superpotential is given by
\[ W=Mb^2.\]
There are lots of checks that this spectrum is indeed 
correct~\cite{IntSeib,discrete}, including integrating out a flavor from 
the theory with $F=N-2$ and obtain this branch, continuous and discrete
anomaly matching, and integrating out one more flavor. However, this
is the only known example of this kind, and it would be very interesting
to find more confining theories of this sort.

\section{Theories with a Vanishing Confining Superpotential}

There are two broad classes of known confining theories with vanishing
superpotential. One class includes the famous theories with a 
quantum deformed moduli space, while the other class contains
the theories 
with an ``affine moduli space'' of vacua. These can be distinguished by 
noting, that in the first case there are non-trivial classical constraints
among the basic composite invariants, while in the second case there are
none. 

\subsection{Theories with Constraints: Quantum Deformed Moduli Space}

The first example of a theory with a quantum modified constraint has 
been discovered by Seiberg~\cite{Seiberg}. The example is SUSY QCD with
the number of colors equal to the number of flavors,
$SU(N)$ with $F=N$. 
The field content and global symmetries of the theory, together with the
confining spectrum is given below.
\[
\begin{array}{c|c|cccc}
& SU(N) & SU(N) & SU(N) & U(1)_B & U(1)_R \\ \hline
Q & \Yfund & \Yfund & 1 & 1 & 0 \\
\bar{Q} & \overline{\Yfund} & 1 & \Yfund & -1 & 0 \\ \hline \hline
M=(\bar{Q}Q) & & \Yfund & \Yfund & 0 & 0 \\
B=(Q^N) &  & 1 & 1 & N & 0 \\
\bar{B} =(\bar{Q}^N) & & 1 & 1 & -N & 0
\end{array} \]
The composites $M, B$ and $\bar{B}$ satisfy the classical constraint
${\rm det}\; M-B\bar{B}=0$. In the infrared, quantum effects modify 
this constraint to ${\rm det}\; M-B\bar{B}=\Lambda^{2N}$. Again there is
a lot of evidence that this is indeed what happens. 
The 't Hooft anomaly matching conditions
 are not satisfied at the origin, but they are satisfied at any point on the
quantum deformed moduli space (from which the origin is excluded).
Integrating out one flavor reproduces the well-known Affleck-Dine-Seiberg
superpotential~\cite{ADS}, and higgsing the gauge group will also give a
consistent result. Later more theories with a quantum modified constraint
have been identified~\cite{Pouliot,Sp,PT}.

\begin{table}
\begin{center}
\caption{The $SU$ theories satisfying the index constraint $\sum_i \mu_i 
=G$. The 
first column gives the gauge group, the second column the field content and
the third column gives the phase of the low-energy theory. QDMS stands for 
confining with a quantum deformed moduli space, and i and c distinguish
between the cases where the constraint which is quantum modified is
invariant or covariant under the global symmetries of the 
theory~\protect\cite{GN}. Note, that all theories satisfying  
$\sum_i \mu_i =G$
are either confining with a quantum modified constraint or in the
Coulomb phase. The theories in the Coulomb phase have been discussed in
Ref.~\protect\cite{CS}.}
\label{tab:qdms}
\[
\begin{array}{|c|c|c|} \hline
SU(N)   &   N (\Yfund + \overline{\Yfund})   & \mbox{i-QDMS}   \\
SU(N)   &  \Yasymm + (N-1)\, \overline{\Yfund} + 3\, \Yfund 
                                              & \mbox{i-QDMS} \\
SU(N) & \Yasymm + \overline{\Yasymm} + 2 (\Yfund +
\overline{\Yfund})     		& \mbox{i-QDMS} \\
SU(N) & Adj   & \mbox{Coulomb phase} \\ 
SU(4) & 3\, \Yasymm +(\Yfund + \overline{\Yfund}) &  \mbox{c-QDMS} \\
SU(4) &  4\, \Yasymm  &   \mbox{Coulomb phase} \\ 
SU(5) &  2\, \Yasymm +  \Yfund + 3\, \overline{\Yfund} 	
					& \mbox{i-QDMS} \\
SU(5) & 2\, \Yasymm + \overline{\Yasymm} +  
	\overline{\Yfund} & \mbox{c-QDMS} \\ 
SU(6) & 2\, \Yasymm + 4\, \overline{\Yfund} & \mbox{i-QDMS} \\
SU(6) & \Ythreea + 3 (\Yfund + \overline{\Yfund}) & \mbox{i-QDMS} \\
SU(6) & \Ythreea + \Yasymm +  2\, \overline{\Yfund} 
& \mbox{c-QDMS} \\
SU(6) &  2\, \Ythreea  & \mbox{Coulomb phase}  \\  
SU(7) &  \Ythreea + 4\, \overline{\Yfund} + 2\, 
		\Yfund  & \mbox{c-QDMS} \\  \hline
\end{array}\]
\end{center}
\end{table}

One can again try to find all theories that similarly have a quantum modified
constraint. In these theories a classical constraint of the form $\sum (\Pi_i X_i)=0$
(where $X_i$ are gauge invariant operators) is modified quantum mechanically
to $\sum (\Pi_i X_i)=\Lambda^p \Pi_j X_j$. Here, the $X_j$ are some other
combination of the gauge invariant operators, including the possibility that the
quantum modification is just $\Lambda^p$. The power $p$ must necessarily be
positive to reproduce the correct classical limit. Such a modification of the
classical constraint is only possible in theories where $\sum \mu_i-G=0$.
To show this, consider assigning R-charge zero to every chiral superfield.
This R-symmetry is anomalous and the anomaly has to be compensated by
assigning R-charge $\sum \mu_i-G$ to the scale of the gauge group
raised to the power of its one loop $\beta$ function coefficient
$\Lambda^{(3 G - \sum \mu_i)/2}$.
Since the constraints have to respect this R-symmetry one immediately
sees that $\Lambda$ can only appear in a constraint if it has vanishing
R-charge. Therefore, we conclude that only theories with
$\sum \mu_i-G=0$ may exhibit quantum deformed moduli spaces.
We can find all theories satisfying $\sum \mu_i-G=0$ by simply leaving out
a flavor from the matter contents listed in Table~\ref{tab:s-conf}. The
resulting theories are displayed in Table~\ref{tab:qdms}. The theories based
on $SU$ groups of Table~\ref{tab:s-conf} have been examined in detail
by Grinstein and Nolte~\cite{GN}, and those based on other groups 
by Grinstein and Nolte~\cite{GN2} and Cho~\cite{Cho}. Again, based on the flows
one can exclude all theories from Table~\ref{tab:s-conf} which do not flow
to a confining theory. In the remaining examples one can find the
quantum modified constraint either by integrating out one flavor from
an s-confining theory, or if the theory with one more flavor is not
s-confining, then one has to consider the flows along various flat directions
in order to find what the quantum modified constraint is. It has been found
in~\cite{GN,GN2} that there are two types of theories with a quantum modified
constraint. One possibility is that the constraint is invariant under all
global symmetries, then the quantum modified constraint has the form
$\Pi_i X_i=\Lambda^p$, and the origin is excluded from the moduli space by
the quantum modification. The other possibility is that the constraint
carries a non-vanishing global charge, and thus the quantum modification must
be field dependent, of the form $\Pi_i X_i=\Lambda^p X_1$, where $X_1$ is
a single composite field. In this case, the origin of the moduli space is
not excluded, and the 't Hooft anomaly matching conditions have to be 
satisfied after the field $X_1$ is eliminated from the spectrum.
Below, we present an example where the quantum modified constraint is
covariant under the global symmetries. The example is based on an $SU(4)$ 
gauge theory with matter in $3\, \Yasymm +\Yfund
+\overline{\Yfund}$. The theory with an additional flavor is not s-confining.
The confining
spectrum is given in the table below.

\begin{displaymath}
\begin{array}{c|c|cccc}
   & SU(4) & SU(3) & U(1)_1 & U(1)_2 & U(1)_R \\ \hline
A  & \Yasymm & \Yfund & 0 & 1 & 0 \\
Q  & \Yfund & 1 & 1 & -3 & 0 \\
\bar{Q} & \overline{\Yfund} & 1 & -1 & -3 & 0 \\ \hline \hline
A^2 & & \Ysymm & 0 & 2 & 0 \\
Q A^2 \bar{Q} & & \overline{\Yfund} & 0 & -4 & 0 \\
Q \bar{Q} & & 1 & 0 & -6 & 0 \\
A^3 Q^2 & & 1 & 2 & -3 & 0 \\
A^3 \bar{Q}^2 & & 1 & -2 & -3 & 0
\end{array}
\end{displaymath}
The quantum modified constraint is
\begin{displaymath}
\frac{1}{6}(Q\bar{Q})^2 (A^2)^3 + 
4(A^2) (Q A^2 \bar{Q})^2 + 64 (A^3 Q^2)(A^3 \bar{Q}^2)^2 = 
\Lambda^8 (Q\bar{Q}) 
\end{displaymath}
Note that one can eliminate the field $(Q\bar{Q})$ from the theory by solving
the quantum modified constraint. The remaining fields match all anomalies 
of the ultraviolet theory.

Finally, we note that there are no known confining theories, where 
classical constraints among the basic invariants do exist, but none of them 
is quantum modified. However, there is no argument why theories like that 
could not exist. 

\subsection{No Constraints Among Invariants: Affine Moduli Space}

The first and perhaps most famous example of this class of theories is
the ISS model~\cite{ISS}, which is an $SU(2)$ gauge theory with one chiral
superfield $Q$ in the spin 3/2 representation of the gauge group. Classically 
this theory has a single independent gauge invariant
$Q^4$, which satisfies the 't Hooft anomaly matching conditions. Therefore
it is widely believed that this theory confines without generating a 
confining superpotential. Theories which have at least a branch on 
which they behave analogously have been later found in 
Refs.~\cite{IntSeib,Sp,CSS}. The classification of such theories has 
been done by Dotti and Manohar. They obtain a list of all theories 
where there is no constraint among the fundamental composites (which they
call theories with an affine moduli space), and on these theories
they explicitly check whether the 't Hooft anomaly matching conditions 
are satisfied or not. The resulting theories are given in Table~\ref{tab:aff}.
The first six theories in Table~\ref{tab:aff} have a confining branch 
with no superpotential generated in addition to a branch with a 
dynamically generated superpotential. The seventh theory is the 
ISS model which as explained above presumably only has a confining phase
with no superpotential generated. The phase of the last four theories is
not very well established. The fact that the 't Hooft anomaly matching 
conditions are satisfied would suggest that these theories are confining just
like the ISS model. However, a more careful analysis of the different branches
of these theories shows that it is unlikely that these theories confine
at the origin, instead they are likely to be in an interacting 
non-Abelian Coulomb phase~\cite{BCI}. 

\begin{table}
\caption{The theories which have no classical constraints among the 
basic invariants and satisfy 't Hooft anomaly matching from 
Ref.~\protect\cite{DM}.
The first column gives the gauge group, the second column the matter content.
$S$ stands for the spinor of the given $SO$ group.}
\label{tab:aff}
\[ 
\begin{array}{|c|c|} \hline 
SU(2N) & \Yasymm + \overline{\Yasymm}   \\
SU(6) & \Ythreea  \\
Sp(2N), N \geq 2 & \Yasymm \\
SO(N), N \geq 5 & (N-4)\Yfund  \\
SO(12) & 2 S  \\
SO(14)& S  \\
SU(2) & \Ythrees  \\
SU(8) & \Yfoura  \\ & \\
Sp(8) & \Yfoura  \\
SO(N), N \geq 5 & \Ysymm  \\
SO(16)& S  \\
\hline
\end{array}\]
\end{table}

Finally, we note that Dotti and Manohar have also shown that the
only theories with no classical invariants at all (which are believed 
to break supersymmetry dynamically) are the two well-known examples:
$SU(5)$ with $10+\bar{5}$ and $SO(10)$ with a single spinor.

\section{Conclusions}

There have been a lot of new results recently concerning the low-energy
dynamics of $N=1$ supersymmetric gauge theories. The simplest of these
theories are the confining ones, where the low-energy effective theory is
simply a theory of gauge singlets. The known confining theories can be 
classified according to whether or not there is a superpotential
generated for the confined degrees of freedom. The theories which do
have a confining superpotential include the s-confining theories and the
theories where the composites involve the gauge field $W_{\alpha}$.
The class of theories where there is no superpotential for the confined 
degrees of freedom contains the theories with a quantum deformed moduli space
and the theories with an affine quantum moduli space. Some of these categories
(s-confining, quantum deformed moduli space, affine moduli space) have 
been exhaustively studied for the case of simple gauge group and no tree-level 
superpotential. Others are not well understood, and perhaps there might
be completely new types of confining theories waiting to be discovered.

\section*{Acknowledgments}
I thank Martin Schmaltz and Witold Skiba for our collaboration 
on Ref.~\cite{CSS},
based on which parts of this review have been written. 
I also thank John Terning for comments on the manuscript.
The author is a 
research fellow of the Miller Institute for Basic Research in Science.
This work was supported in part by the U.S. 
Department of Energy under Contract DE-AC03-76SF00098, and in part by the 
National Science Foundation under grant PHY-95-14797.


\section*{References}
\begin{thebibliography}{99}

\bibitem{CSS} C. Cs\'aki, M. Schmaltz and W. Skiba, \PRL{78}{799}{1997}, 
hep-th/9610139;  \PRD{55}{7840}{1997}, hep-th/9612207. 

\bibitem{Seiberg} N. Seiberg, \PRD{49}{6857}{1994}, hep-th/9402044. 

\bibitem{Pouliot} K. Intriligator and P. Pouliot, \PLB{353}{471}{1995}, 
hep-th/9505006; 
P. Pouliot, \PLB{367}{151}{1996}, hep-th/9510148;
\PLB{359}{108}{1995}, hep-th/9507018; I. Pesando, \MPLA{10}{1871}{1995}, 
hep-th/9506139; S. Giddings and J. Pierre,  \PRD{52}{6065}{1995}, 
hep-th/9506196. 


\bibitem{Sp} P. Cho and P. Kraus, \PRD{54}{7640}{1996}, 
hep-th/9607200; C. Cs\'aki, M. Schmaltz and W. Skiba,
\NPB{487}{128}{1997}, hep-th/9607210.

\bibitem{IntSeib} K. Intriligator and N. Seiberg, \NPB{444}{125}{1995}, 
hep-th/9503179. 

\bibitem{discrete} C. Cs\'aki and H. Murayama, \NPB{515}{114}{1998}, 
hep-th/9710105. 

\bibitem{ADS} I. Affleck, M. Dine and N. Seiberg, \NPB{241}{493}{1984}.  

\bibitem{PT} E. Poppitz and S. Trivedi, \PLB{365}{125}{1996}, 
hep-th/9507169. 

\bibitem{GN} B. Grinstein and D. Nolte, \PRD{57}{6471}{1998}, 
hep-th/9710001. 

\bibitem{GN2} B. Grinstein and D. Nolte,  hep-th/9803139. 

\bibitem{Cho} P. Cho, \PRD{57}{5214}{1998}, 
hep-th/9712116.

\bibitem{CS} C. Cs\'aki and W. Skiba,
\PRD{58}{045008}{1998}, hep-th/9801173.
 
\bibitem{ISS} K. Intriligator, N. Seiberg and S. Shenker, \PLB{342}{152}{1995},
hep-ph/9410203. 


 
\bibitem{DM} G. Dotti and A. Manohar, \PRL{80}{2758}{1998}, 
hep-th/9712010. 

\bibitem{BCI} J. Brodie, P. Cho and K. Intriligator, \PLB{429}{319}{1998}, 
hep-th/9802092. 

\end{thebibliography}
\end{document}